\documentclass{article}
\usepackage{spconf,amsmath,graphicx}
\usepackage{times}
\usepackage{epsfig}
\usepackage{hhline}
\usepackage{amsmath}
\usepackage{amssymb}
\usepackage{algorithm}
\usepackage{algorithmic}
\usepackage{enumerate}
\usepackage{multirow}
\usepackage{multicol}
\usepackage{arydshln}
\usepackage{color}
\usepackage{amssymb}
\usepackage{graphicx}
\usepackage{subfigure}
\usepackage{rotating}
\usepackage{multirow}
\usepackage[table,xcdraw]{xcolor}
\def\0{{\mathbf 0}}
\def\1{{\mathbf 1}}

\def\b{{\mathbf b}}
\def\c{{\mathbf c}}

\def\g{{\mathbf g}}

\def\l{{\mathbf l}}

\def\p{{\mathbf p}}

\def\r{{\mathbf r}}

\def\x{{\mathbf x}}
\def\y{{\mathbf y}}
\def\z{{\mathbf z}}

\def\A{{\mathbf A}}

\def\F{{\mathbf F}}
\def\G{{\mathbf G}}
\def\F{{\mathbf F}}
\def\H{{\mathbf H}}
\def\I{{\mathbf I}}

\def\L{{\mathbf L}}

\def\P{{\mathbf P}}

\def\W{{\mathbf W}}

\def\ie{{\textit{i.e.}}}
\def\eg{{\textit{e.g.}}}

\def\cE{{\mathcal E}}

\def\cG{{\mathcal G}}

\def\cL{{\mathcal L}}

\def\cN{{\mathcal N}}
\def\cO{{\mathcal O}}

\title{Retinex-based Image Denoising / Contrast Enhancement using \\ Gradient Graph Laplacian Regularizer}
\name{Yeganeh Gharedaghi$^\dag$, Gene Cheung$^\dag$, Xianming Liu$^\ddag$\thanks{The work of G. Cheung was supported in part by the Natural Sciences and Engineering Research Council of Canada (NSERC) RGPIN-2019-06271, RGPAS-2019-00110.}}
\address{$^\dag$York University, Toronto, Canada ~~~~~~~ $^\ddag$Harbin Institute of Technology, Harbin, China}
\begin{document}
\ninept
\maketitle
\begin{abstract}
Images captured in poorly lit conditions are often corrupted by acquisition noise.  
Leveraging recent advances in graph-based regularization, we propose a fast Retinex-based restoration scheme that denoises and contrast-enhances an image. 
Specifically, by Retinex theory we first assume that each image pixel is a multiplication of its reflectance and illumination components. 
We next assume that the reflectance and illumination components are piecewise constant (PWC) and continuous piecewise planar (PWP) signals, which can be recovered via graph Laplacian regularizer (GLR) and gradient graph Laplacian regularizer (GGLR) respectively. 
We formulate quadratic objectives regularized by GLR and GGLR, which are minimized alternately until convergence by solving linear systems---with improved condition numbers via proposed preconditioners---via conjugate gradient (CG) efficiently.
Experimental results show that our algorithm achieves competitive visual image quality while reducing computation complexity noticeably.
\end{abstract}
\section{Introduction}
\label{sec:intro}
Due to the relatively few photons collected per pixel area, a sensor capturing an image in poor lighting conditions suffers from non-negligible acquisition noise.
Thus, a contrast enhancement algorithm such as \cite{DBLP:conf/cvpr/FuZHZD16} that selectively brightens spatial areas to produce a visually pleasing image would also enhance the acquired noise, resulting in sub-par image quality.
We study the joint image denoising / contrast enhancement problem in this paper.

Since Land's seminal Retinex theory in human vision in 1977
\cite{land77}, researchers in imaging have since interpreted the theory to mean that a recorded pixel is a multiplication of \textit{illumination} and \textit{reflectance} components 
\cite{kimmel03,ma11}. 
Because the two components have unique signal characteristics---\eg, reflectance is commonly assumed to be \textit{piecewise constant} (PWC)---image restoration schemes can be designed to first recover these components with appropriate signal priors, before combining them to reconstruct the target image \cite{kimmel03,ma11,DBLP:journals/tip/GuoLL17,DBLP:journals/corr/abs-1804-08468,ren20,yao21,lecert22}. 
However, computation of the illumination and/or reflectance components can be expensive; for example, \cite{ma11} employed Bergman iteration to minimize an $\ell_1$-norm objective, while \cite{ren20} proposed a non-convex low-rank signal prior. 
Deep learning based Retinex schemes are also possible \cite{chen18,li22}, but they require expensive data training for a large number of network parameters with a large memory footprint, and thus are not typically suitable for memory-constrained devices like mobile phones.

In an orthogonal development, \textit{graph signal processing} (GSP) has been intensively investigated over the last decade to study discrete signals on irregular data kernels described by graphs \cite{ortega18ieee,cheung18}. 
In restoration problems, graph-based regularization terms like \textit{graph Laplacian regularizers} (GLR) \cite{pang17} have been adopted for a wide range of applications, including joint image contrast enhancement / JPEG dequantization in \cite{liu19}.  
Like total variation (TV) \cite{chambolle97}, \textit{signal-dependent} GLR (SDGLR) has been shown to promote PWC signal reconstruction \cite{pang17,liu17}, but unlike non-differentiable $\ell_1$ norm, GLR is in differentiable quadratic form that is amenable to fast optimization. 
\cite{liu19} employed GLR to efficiently recover the reflectance component via proximal gradient descent \cite{parikh13}. 

In this paper, leveraging \cite{liu19} we employ a new variant of GLR called \textit{gradient graph Laplacian regularizer} (GGLR) \cite{chen22}---shown to promote continuous \textit{piecewise planar} (PWP) signal reconstruction---to recover the illumination component known to be generally smooth.
Like GLR, GGLR is also in convenient quadratic form, leading to a system of linear equations for a solution computed efficiently using \textit{conjugate gradient} (CG) \cite{shewchuk94}. 
Moreover, we propose appropriate \textit{preconditioners} \cite{golub12} to improve the condition numbers of the coefficient matrices, speeding up CG execution.
We leave the unrolling of our iterative graph-based algorithm to neural layers for data-driven end-to-end parameter optimization \cite{vu21} for future work.
After recovering the reflectance and illumination components, the contrast-enhanced image is reconstructed via gamma correction \cite{gonzalez2008digital} on the illumination component.
Experimental results show that our method has comparable contrast-enhanced image quality as competing schemes with reduced computation costs. 

\section{Preliminaries}
\label{sec:prelim}
\subsection{GSP Basics}

We first review GSP definitions \cite{ortega18ieee}. 
A graph $\cG(\cN,\cE,\W)$ is composed of $N$ nodes $\cN = \{1, \ldots, N\}$ and edges $\cE$ connecting them, where edge $(i,j) \in \cE$ has weight $w_{i,j} = W_{i,j}$.
Assuming that edges are undirected, the \textit{adjacency matrix} $\W$ is symmetric.
The combinatorial \textit{graph Laplacian matrix} is defined as $\L \triangleq \text{diag}(\W \1) - \W$, where $\1$ is an all-one vector of suitable length, and $\text{diag}(\W \1)$ denotes a diagonal matrix with vector $\W \1$ as diagonal terms. 
$\L$ is provably \textit{positive semi-definite} (PSD)---\ie, $\x^\top \L \x \geq 0, \forall \x$---if edges are non-negative, \ie, $w_{i,j} \geq 0, \forall i,j$ \cite{cheung18}. 

$\x^\top \L \x$ is also called the \textit{graph Laplacian regularizer} (GLR), and its signal-dependent variant---where each edge weight $w_{i,j}$ is a function of sought signal samples $x_i$ and $x_j$---has been shown to promote PWC signal reconstruction \cite{pang17,liu17}.
It was used for regularization in different graph signal restoration problems, including image denoising \cite{pang17}, JPEG dequantization \cite{liu17}, point cloud denoising \cite{zeng20,dinesh20} and super-resolution \cite{dinesh22}.
Other graph-based regularizations are possible, such as \textit{graph total variation} (GTV) \cite{bai19} and \textit{graph shift varation} (GSV) \cite{chen15}. 
In this work, we focus on GLR and a recent variant called \textit{gradient graph Laplacian regularizer} (GGLR), which was shown to promote PWP signal reconstruction \cite{chen22}. 
For images, GGLR means applying GLR to horizontal / vertical image gradients; we detail derivation of GGLR in Section\;\ref{subsec:reflectance}.

\subsection{Interpretation of Retinex Theory}

Similar to \cite{kimmel03,ma11,liu19}, we mathematically interpret the known Retinex theory \cite{land77} to mean that a ground-truth $N$-by-$N$ image patch (vectorized to strictly positive $\x \in \mathbb{R}_+^{N^2}$ by scanning pixels row-by-row) is a point-by-point multiplication of strictly positive illumination and reflectance components, $\l, \r \in \mathbb{R}_+^{N^2}$, \ie, $\x = \l \odot  \r$, where operator $\odot$ denotes point-by-point multiplication. 
Specifically, the image formation model for observation $\y \in \mathbb{R}_+^{N^2}$ is 
\begin{align}
\y = \l \odot \r + \z ,
\label{eq:retinex}
\end{align}
where 
$\z$ is a zero-mean additive Gaussian noise.
Reflectance $\r$ depends only on surface properties of physical objects, and is known to be piecewise smooth or PWC. 
In contrast, illumination $\l$ varies less drastically than $\r$; we model $\l$ to be continuous PWP in this work.

\section{Problem Formulation}
\label{sec:formulate}
\subsection{Initialization of Illumination \& Reflectance}

Our method requires initialization of both the illumination and reflectance components, $\l$ and $\r$, of a $N$-by-$N$ pixel patch, after which one component is optimized while the other is held fixed. 
To initialize $\l$, we compute the blurred V component of an input image in the HSV color space using a Gaussian filter with a standard deviation of $5$. 
We then initialize $\r$ by performing a point-by-point division of the image's intensity component ($\y$) by $\l$.

\subsection{Computation of Reflectance} \label{subsec:reflectance}

\subsubsection{Graph Construction}

To connect a group of $N$ reflectance pixels in the $k$-th row ($k$-th column) of a target $N$-by-$N$ pixel patch, we construct a graph $\cG_{r,k}$ ($\cG_{c,k}$) as follows.
We connect pixels $i$ and $j$ in row $k$ (column $k$) with edge weight $w_{i,j}^{r,k}$ ($w_{i,j}^{c,k}$) defined as
\begin{align}
w_{i,j}^{r,k} = \exp \left( - \frac{|r_i - r_j|^2}{\sigma_r^2} 
- \frac{\|\c_i - \c_j\|^2_2}{\sigma_c^2} 
\right) ,
\label{eq:edge_reflectance}
\end{align}
where $r_i$ and $\c_i$ are the reflectance intensity and 2D-coordinate of pixel $i$, respectively, and $\sigma_r$ and $\sigma_c$ are two parameters.
\eqref{eq:edge_reflectance} is analogous to \textit{bilateral filter} weights \cite{tomasi98}, where the 2D-coordinates compute the \textit{domain filter}, and the reflectance values compute the \textit{range filter}. 
Note that edge weights in \eqref{eq:edge_reflectance} are \textit{signal-dependent}---edge weights $\{w^{r,k}_{i,j}\}$ used to compute $\r$ depend on $\r$.
For a sparse graph, $w_{i,j}^{r,k}$ exists iff $j$ is in a local neighborhood $\cN_i$ of pixel $i$.
See Fig.\;\ref{fig:graphEx}(a) for an example of a line graph for a row of three pixels.

Collection of edge weights $\{w_{i,j}^{r,k}\}$ \eqref{eq:edge_reflectance} defines a symmetric \textit{adjacency matrix} $\W_{r,k} \in \mathbb{R}^{N \times N}$, and the corresponding \textit{graph Laplacian matrix} is defined as $\L_{r,k} \triangleq \text{diag}(\W_{r,k} \1) - \W_{r,k} \in \mathbb{R}^{N \times N}$.
As discussed, $\L_{r,k}$ is PSD given non-negative edges in \eqref{eq:edge_reflectance}.
We use notations $\W_{c,k}$ and $\L_{c,k}$ for the adjacency and graph Laplacian matrices of graph $\cG_{c,k}$ for the $k$-th column of a target patch.

\subsubsection{Optimizing Reflectance}

Given illumination $\l$ for a $N^2$-pixel patch, we compute reflectance $\r$ by minimizing an unconstrained convex quadratic objective:
\begin{align}
\min_{\r} ~& \|\y - \text{diag}(\l) \r \|^2_2 + \mu_r \sum_{k=1}^N \r^\top \H_k ^\top \L_{r,k} \H_k \r
\nonumber \\
&+ \mu_r \sum_{k=1}^N \r^\top \G_k ^\top \L_{c,k} \G_k \r
\label{eq:opt_reflectance}
\end{align}
where $\H_k, \G_k \in \{0,1\}^{N \times N^2}$ are \textit{selection matrices} that pick out $N$ pixels from the $k$-th row / column of the target patch, respectively. 
For example, $\H_1$ and $\G_2$ for a $2 \times 2$ patch are
\begin{align}
\H_1 = \left[ \begin{array}{cccc}
1 & 0 & 0 & 0 \\
0 & 1 & 0 & 0
\end{array} \right], ~~
\G_2 = \left[ \begin{array}{cccc}
0 & 1 & 0 & 0 \\
0 & 0 & 0 & 1
\end{array} \right] .
\end{align}

In \eqref{eq:opt_reflectance}, the first term is a fidelity term following the image formation model \eqref{eq:retinex}, and the second and third terms are \textit{graph Laplacian regularizers} (GLR) \cite{pang17} for the rows and columns of $\r$, respectively. 
$\mu_r$ is a parameter that trades off the fidelity term and the GLRs. 
Selecting rows and columns from a two-dimensional pixel grid for regularization means smaller Laplacian matrices, and thus lower complexity. 
Moreover, promoting a piecewise linear (constant) 1D signal across each dimension separately can combine to mean promotion of a piecewise planar (constant) signal on a 2D grid.

The solution $\r^*$ to \eqref{eq:opt_reflectance} is obtained by solving a linear system:

\vspace{-0.05in}
\begin{footnotesize}
\begin{align}
\left( \text{diag}^2(\l) + \mu_r \sum_{k=1}^N (\H_k^\top \L_{r,k} \H_k + \G_k^\top \L_{c,k} \G_k) \right) \r^* = \text{diag}(\l) \y . 
\label{eq:sol_reflectance}
\end{align}
\end{footnotesize}\noindent 
\eqref{eq:sol_reflectance} guarantees a unique solution because the coefficient matrix $\A = \text{diag}^2(\l) + \mu_r \sum_k (\H_k^\top \L_{r,k} \H_k + \G_k^\top \L_{c,k} \G_k)$ is provably \textit{positive definite} (PD): $\text{diag}^2(\l)$ is PD and $\{\L_{r,k}\}$ and $\{\L_{c,k}\}$ are PSD, and thus $\A$ is PD by Weyl's inequality \cite{golub12}.
Given that coefficient matrix $\A$ is sparse, symmetric and PD, $\r^*$ in \eqref{eq:sol_reflectance} can be computed via \textit{conjugate gradient} (CG) \cite{shewchuk94} in roughly linear time without matrix inversion. 
We defer discussion of complexity to Section\;\ref{subsec:complexity}.

\subsection{Computation of Illumination}

Computation of the illumination component $\l$ differs from reflectance in that we assume $\l$ is generally smooth instead of PWC; mathematically we regularize $\l$ using GGLR \cite{chen22}. 
This requires first the construction of a \textit{gradient graph} for each row / column of pixels in a target image patch, on which we define a GLR.
Then we map the gradient GLR back to the pixel domain as GGLR for optimization.

\subsubsection{Graph Construction}

\begin{figure}[t]
\begin{center}
\includegraphics[width=0.9\linewidth]{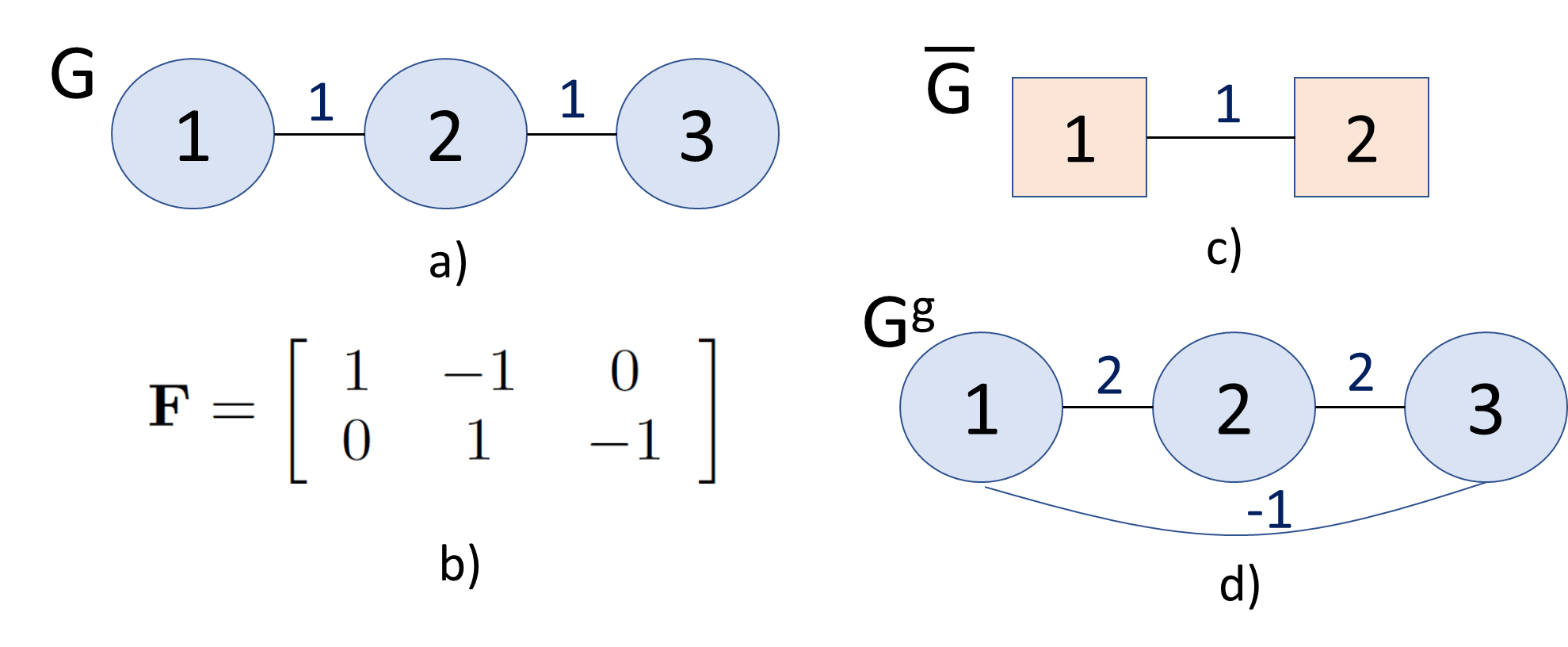}
\end{center}
\vspace{-0.3in}
\caption{A 3-node line graph $\cG$ in (a), a gradient operator $\F$ for a row / column of 3 pixels in (b), gradient graph $\bar{\cG}$ in (c), and resulting GNG $\cG^g$ in (d)---a signed graph with positive / negative edges. } 
\label{fig:graphEx}
\end{figure}

For a $k$-th row (column) of $N$ pixels in a $N^2$ pixel patch, \ie, $\H_k \l$ ($\G_k \l$), we first define a \textit{gradient operator} $\F \in \mathbb{R}^{N-1 \times N}$ as
\begin{align}
F_{i,j} &= \left\{ \begin{array}{ll}
1 & \mbox{if} ~ i=j \\
-1 & \mbox{if} ~ i=j+1 \\
0 & \mbox{o.w.}
\end{array} \right. .
\end{align}
See Fig.\;\ref{fig:graphEx}(b) for an example of gradient operator $\F$ for a row of three pixels. 
Note that $\F \1 = \0$, and $\F$ is full row-rank \cite{chen22}. 
Given $\H_k \l$, we first compute horizontal gradient $\g = \F \H_k \l \in \mathbb{R}^{N-1}$. 
We then construct a \textit{gradient graph} $\bar{\cG}_{r,l}$ to connect $N-1$ gradients in $\g$ as follows.
We connect gradients $i$ and $j$ with edge weight $\bar{w}_{i,j}^{r,k}$:
\begin{align}
\bar{w}_{i,j}^{r,k} = \exp \left( - \frac{|g_i - g_j|^2}{\sigma_l^2} - \frac{\|\c_i - \c_j\|^2_2}{\sigma_c^2}   \right) .
\label{eq:edge_illumination}
\end{align}
Collection of edge weights $\{\bar{w}_{i,j}^{r,k}\}$ \eqref{eq:edge_illumination} thus defines adjacency matrix $\bar{\W}_{r,k}$ and subsequent \textit{gradient graph Laplacian} $\bar{\L}_{r,k} \triangleq \text{diag}(\bar{\W}_{r,k} \1) - \bar{\W}_{r,k}$.
Because $\bar{w}_{i,j}^{r,k} \geq 0, ~\forall i,j$, Laplacian $\bar{\L}_{r,k}$ is provably PSD \cite{cheung18}.
We use notations $\bar{\W}_{c,k}$ and $\bar{\L}_{c,k}$ for the adjacency and graph Laplacian matrices of gradient graph $\bar{\cG}_{c,k}$ for the $k$-th column of a target patch.

\subsubsection{Optimizing Illumination} 

Given $\bar{\L}_{r,k}$, we define GLR for gradient $\g$:
\begin{align}
\g^\top \bar{\L}_l \g = \l^\top \H_k^\top \underbrace{\F^\top \bar{\L}_{r,k} \F}_{\cL_{r,k}} \H_k \l ,
\end{align}
where $\cL_{r,k} = \F^\top \bar{\L}_l \F$ is \textit{gradient-induced nodal graph} (GNG) Laplacian.
$\cL_{r,k}$ corresponds to a graph $\cG_{r,k}^g$ connecting $N$ illumination pixels, which in general is a \textit{signed} graph containing both positive and negative edges; see Fig.\;\ref{fig:graphEx}(c) and (d) for an example of a gradient graph $\bar{\cG}$ and the resulting GNG $\cG^g$ corresponding to a three-pixel row. 
Though Laplacians of general signed graphs can be indefinite, because $\bar{\L}_{r,k}$ is PSD, $\cL_{r,k} = \F^\top \bar{\L}_{r,k} \F$ is also PSD.
$\l^\top \H_k^\top \cL_{r,k} \H_k \l$ is the GGLR for the $k$-th pixel row of the target illumination patch.

Given reflectance $\r$, we compute  illumination $\l$ by minimizing the following objective:
\begin{align}
\min_{\l} &~ \|\y - \text{diag}(\r) \l \|^2_2 + \mu_l \sum_{k=1}^N \l^\top \H_k^\top \cL_{r,k} \H_k \l 
\nonumber \\
& + \mu_l \sum_{k=1}^N \l^\top \G_k^\top \cL_{c,k} \G_k \l ,
\label{eq:opt_illumination}
\end{align}
where the first term is a fidelity term given fixed $\r$, and the second and third terms are GGLRs for the rows and columns of $\l$.
$\mu_l$ is a tradeoff parameter like $\mu_r$ in \eqref{eq:opt_reflectance}.
Similarly, the solution $\l^*$ to \eqref{eq:opt_illumination} can be obtained by solving a linear system:

\vspace{-0.05in}
\begin{footnotesize}
\begin{align}
\left( \text{diag}^2(\r) + \mu_l \sum_{k=1}^N (\H_k^\top \cL_{r,k} \H_k + \G_k^\top \cL_{c,k} \G_k) \right) \l^* = \text{diag}(\r) \y .
\label{eq:sol_illumination}
\end{align}
\end{footnotesize}\noindent 
$\l^*$ in \eqref{eq:sol_illumination} again can be obtained via CG without matrix inversion.

\subsection{Algorithm Summary}

Having obtained $\l^*$, a new solution $\r^*$ to \eqref{eq:sol_reflectance} can be computed again using the updated $\l^*$ and new graphs $\{\cG_{r,k}\}$ and $\{\cG_{c,k}\}$ with edges updated via \eqref{eq:edge_reflectance} using the most recently computed $\r$. 
This iterative update of edge weights means the GLRs are signal-dependent and thus promote PWC reflectance reconstruction.
Having obtained $\r^*$, a new solution $\l^*$ to \eqref{eq:sol_illumination} can be sought using recomputed $\r^*$ and new GNGs $\{\cG_{r,k}^g\}$ and $\{\cG_{c,k}^g\}$ with edges updated via \eqref{eq:edge_illumination}.
Similarly, this iterative edge weight update means the GGLRs are signal-dependent and promote PWP illumination reconstruction.

Having obtained solutions $\l^*$ and $\r^*$, we construct a contrast-enhanced pixel $x_{i,j}$ via \textit{gamma correction} \cite{gonzalez2008digital}: 
\begin{align}
x_{i,j} = \left( l_{i,j} \right) ^\gamma \, r_{i,j} ,
\end{align}
where $0 < \gamma <1$ is a pre-chosen parameter. 
The operation $(l_{i,j})^\gamma$ essentially boosts the illumination component---more enhancement when $l_{i,j}$ is small, and less enhancement when $l_{i,j}$ is large.

\subsection{Computation Considerations}
\label{subsec:complexity}

The complexity of CG to solve for $\x$ in a linear system $\A \x = \b$---assuming $\A$ is symmetric and PD---is $\cO(\sqrt{\kappa(\A)} \, \text{nnz}(\A) / \log(\epsilon))$, where $\kappa(\A) \triangleq \frac{\lambda_{\max}(\A)}{\lambda_{\min}(\A)}$ is the \textit{condition number} of $\A$, $\text{nnz}(\A)$ is the number of non-zero entries in $\A$, and $\epsilon$ is a convergence parameter.  
In linear system \eqref{eq:sol_reflectance}, coefficient matrix $\A = \text{diag}^2(\l) + \mu_r \sum_k (\H_k^\top \L_{r,k} \H_k + \G_k^\top \L_{c,k} \G_k)$ is sparse, symmetric and PD, but $\kappa(\A)$ can be large due to small illumination values in $\l$. 
(Similarly, for linear system \eqref{eq:sol_illumination} $\kappa(\A)$ can be large due to small reflectance values in $\r$.)
To improve the computation speed of CG, we perform \textit{preconditioning} \cite{golub12} as follows. 

Generally, in place of linear system $\A \x = \b$, we can consider equivalent $\P \A \x = \P \b$ instead for invertible matrix $\P$: 
\begin{align}
\P \A \P^\top (\P^\top)^{-1} \x &= \P \b    
\nonumber \\
\P \A \P^\top \hat{\x} &= \hat{\b}
\label{eq:precondition}
\end{align}
where $\P^\top (\P^\top)^{-1} = \I$, $\hat{\x} = (\P^\top)^{-1} \x$, and $\hat{\b} = \P \b$. 
Note that coefficient matrix $\P \A \P^\top$ in \eqref{eq:precondition} is also symmetric and PD given $\A$ is symmetric and PD by assumption, and thus CG can solve \eqref{eq:precondition}.
If $\kappa(\P \A \P^\top) \ll \kappa(\A)$, then we have improved the conditioning of the linear system, and CG will run faster.
The challenge is to design an invertible $\P$ satisfying this condition. 

One simple preconditioner that is easily invertible is a diagonal matrix (called \textit{Jacobi} in the linear algebra literature \cite{golub12}).
We propose one variant: $\P = \text{diag}(\p)$, where  $p_i = A_{i,i}^{-1/2}$. 
This means $\P \A \P^\top = \text{diag}(\p) \A \text{diag}(\p)$ has ones along its diagonal. 
Note that $A_{i,i} > 0$ for $\A$ in \eqref{eq:sol_reflectance}, since $\l$ is strictly positive and diagonals of $\H_k^\top \L_{r,k} \H_k$ and $\G_k^\top \L_{c,k} \G_k$ are non-negative. 
Note also that $\A$ in \eqref{eq:sol_reflectance} is \textit{diagonally dominant}, \ie, $A_{i,i} > \sum_{j\neq i} |A_{i,j}|$---a matrix condition where Jacobi preconditioner is known to perform well.  

Thus, when solving for $\r^*$ in \eqref{eq:sol_reflectance}, we first solve for $\hat{\r}^*$ via linear system $\text{diag}(\p) \, \A \, \text{diag}(\p) \, \hat{\r}^* = \text{diag}(\p) \, \b$. 
We then obtain solution $\r^* = \text{diag}(\p) \, \hat{\r}^*$. 
Similar procedure is employed when solving for $\l^*$ in \eqref{eq:sol_illumination}.

\section{Experiments}
\label{sec:results}

\vspace{-0.05in}
\subsection{Experimental Setup}

\vspace{-0.05in}
We conducted experiments using MATLAB R2022b on an Apple M2 chip with 8GB RAM to evaluate the performance of our proposed method. We selected 12 images from the datasets provided in \cite{DBLP:conf/cvpr/FuZHZD16} and \cite{DBLP:journals/tip/GuoLL17}. 
To ensure compatibility with the chosen patch size, we adjusted the size of each image so that the height and width were divisible by 5. 
We added zero-mean Gaussian noise with a standard deviation of $0.001$ to every image pixel. 
Four parameters in our method, $\mu_r$ in \eqref{eq:opt_reflectance}, $\sigma_r$ in \eqref{eq:edge_reflectance}, $\mu_l$ in \eqref{eq:opt_illumination}, and $\sigma_l$ in \eqref{eq:edge_illumination}, were empirically set to $1$, $1$, $0.1$, and $0.2$, respectively. 
More generally, weight parameters $\mu_r$ and $\mu_l$ can be chosen to minimize mean squared error \cite{DBLP:journals/spl/ChenL17}.
We set the convergence tolerance for CG to $\epsilon = 10^{-6}$.

\vspace{-0.1in}
\subsection{Experimental Results}

\vspace{-0.05in}
We compared our method against four competing schemes \cite{DBLP:conf/cvpr/FuZHZD16, DBLP:journals/tip/GuoLL17, DBLP:journals/corr/abs-1804-08468, ren20}. 
The first two methods focused on contrast enhancement, while the latter two employed joint denoising and contrast enhancement. 
Note that \cite{DBLP:journals/tip/GuoLL17} incorporated BM3D \cite{4271520} as a post-denoising step, where we chose $\sigma = 10$.
Towards a fair comparison, we adjusted the brightness parameter in all five methods so that they produced images with roughly the same brightness level. 

\begin{figure}[ht]
  \centering
  \subfigure[Noisy image]{\includegraphics[width=0.48\linewidth]{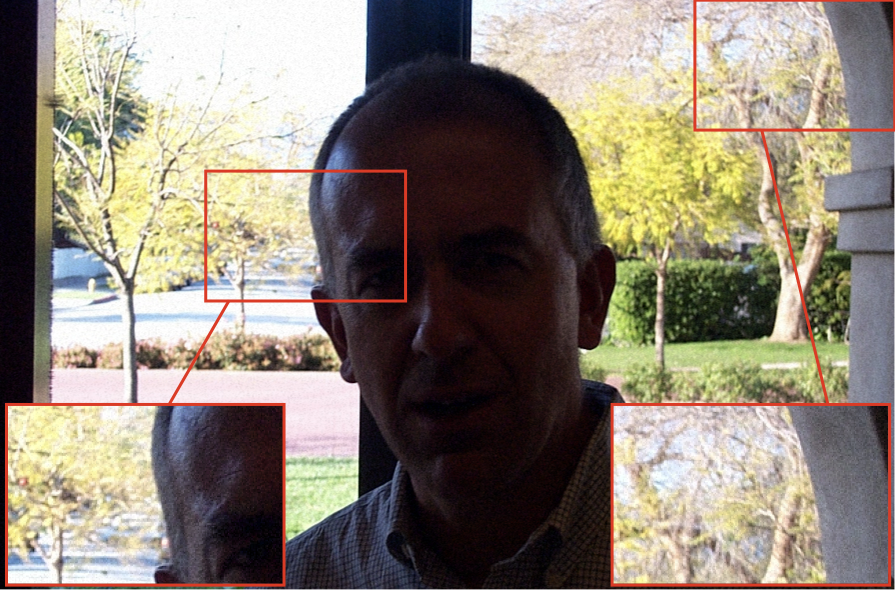}}
  \subfigure[Model from \cite{DBLP:conf/cvpr/FuZHZD16}]{\includegraphics[width=0.48\linewidth]{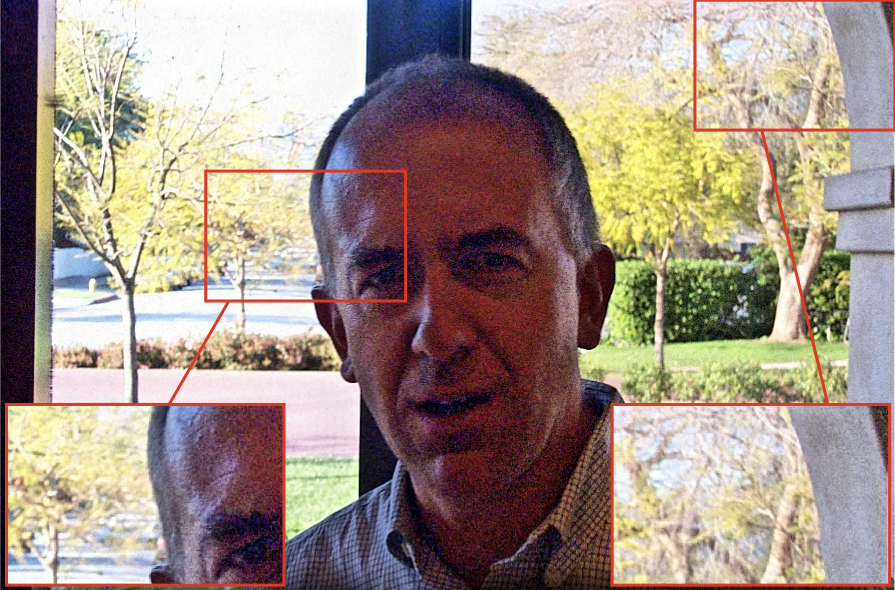}}
  \subfigure[LIME \cite{DBLP:journals/tip/GuoLL17}]{\includegraphics[width=0.48\linewidth]{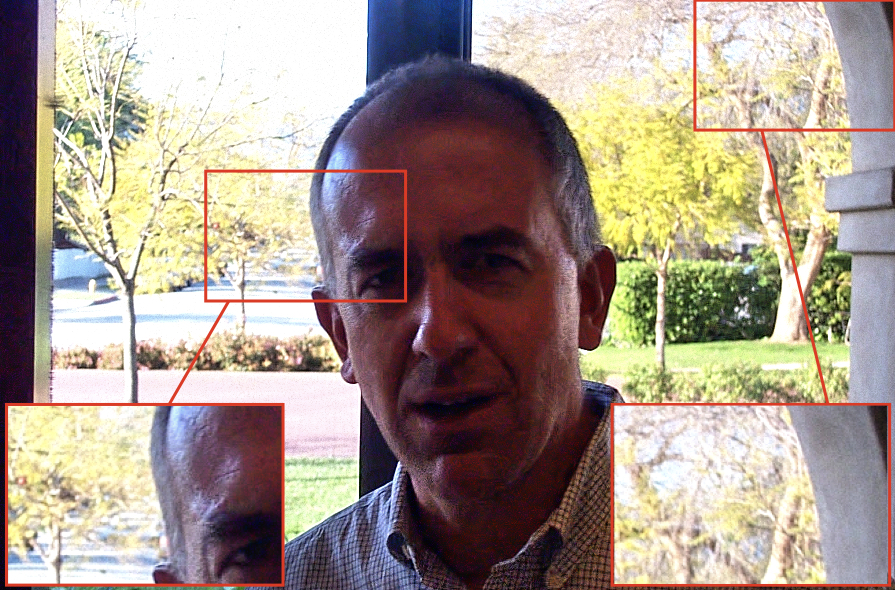}}
  \subfigure[JED \cite{DBLP:journals/corr/abs-1804-08468}]{\includegraphics[width=0.48\linewidth]{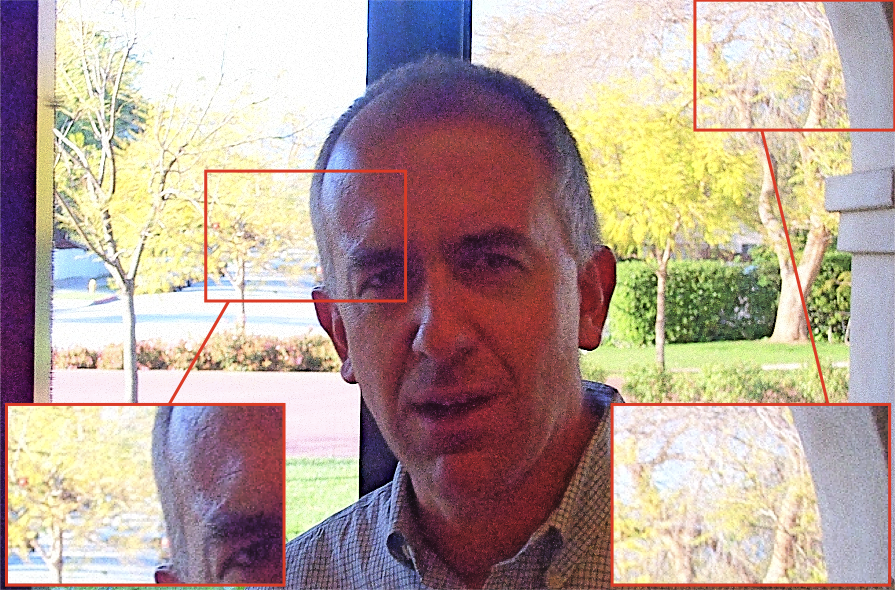}}
  \subfigure[LR3M \cite{ren20}]{\includegraphics[width=0.48\linewidth]{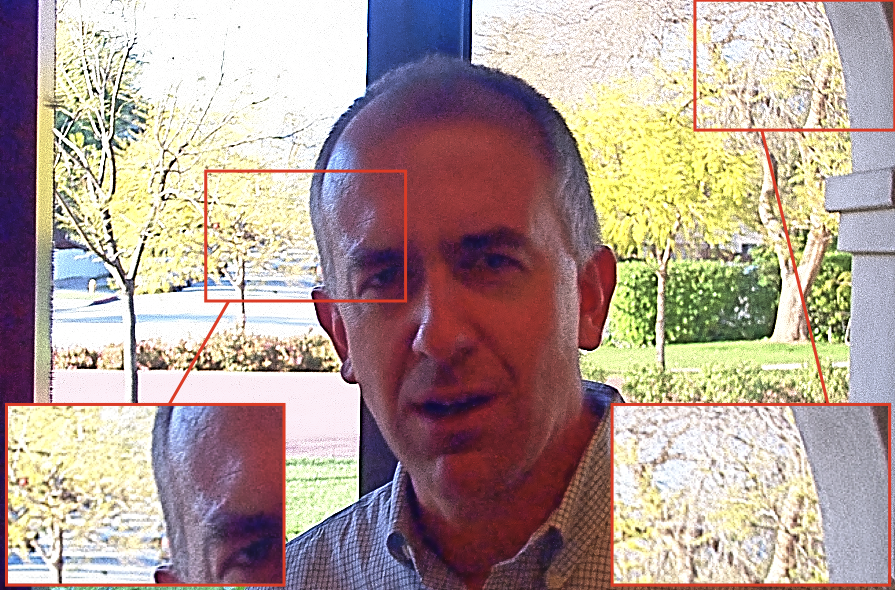}}
  \subfigure[Proposed]{\includegraphics[width=0.48\linewidth]{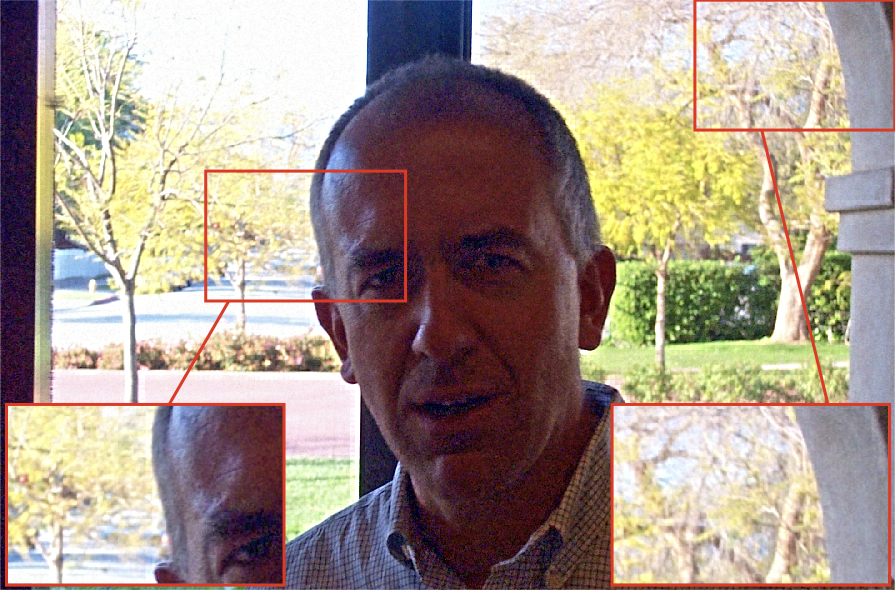}}
  \vspace{-0.15in}
  \caption{Visual comparison of different methods on \texttt{Man}.}
  \label{fig:Man}
\end{figure}
\begin{figure}[ht]
  \centering
  \subfigure[Noisy image]{\includegraphics[width=0.48\linewidth]{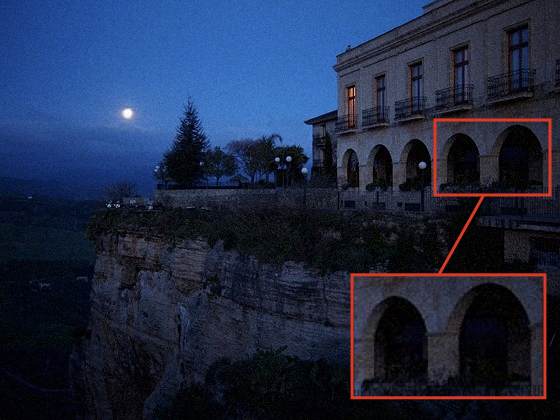}}
  \subfigure[Model from \cite{DBLP:conf/cvpr/FuZHZD16}]{\includegraphics[width=0.48\linewidth]{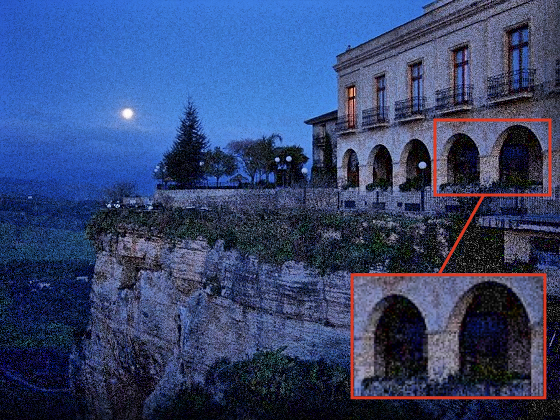}}
  \subfigure[LIME \cite{DBLP:journals/tip/GuoLL17}]{\includegraphics[width=0.48\linewidth]{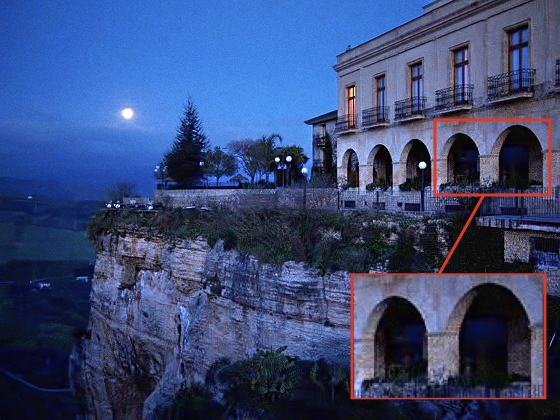}}
  \subfigure[JED \cite{DBLP:journals/corr/abs-1804-08468}]{\includegraphics[width=0.48\linewidth]{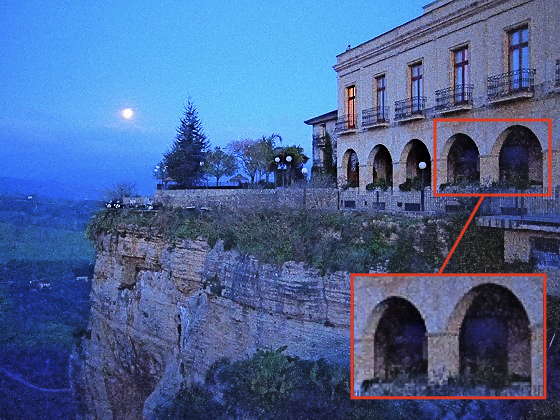}}
  \subfigure[LR3M \cite{ren20}]{\includegraphics[width=0.48\linewidth]{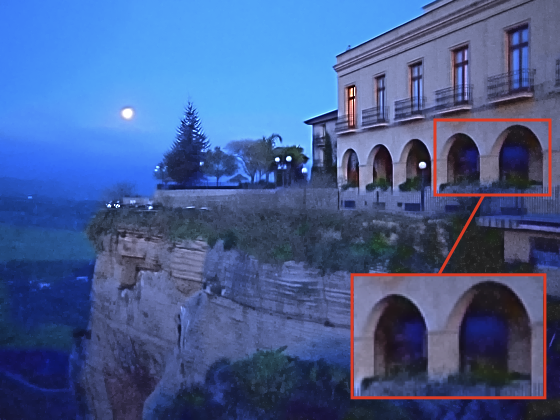}}
  \subfigure[Proposed]{\includegraphics[width=0.48\linewidth]{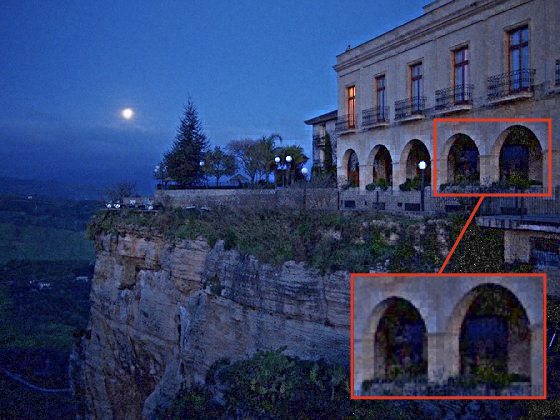}}
  \vspace{-0.15in}
  \caption{Visual comparison of different methods on \texttt{Moonlight}.}
  \label{fig:Moonlight}
\end{figure}

Fig.\;\ref{fig:Man} and \ref{fig:Moonlight} show visual comparisons of different methods. 
The first and third schemes noticeably amplified the noise when performing contrast enhancement, resulting in noisier outputs compared to our method. 
Meanwhile, the second method employed denoising as a post-processing technique, resulting in blurring of image details. 
In comparison, our method produced results that are comparable in quality to the LR3M model with a significantly faster computation speed as demonstrated in Table \ref{tab:speedComparison}. Furthermore, as shown in Fig. \ref{fig:Moonlight}, LR3M can lead to image over-smoothing.

In our objective evaluation, we have examined our method in contrast enhancement using two metrics: Lightness-Order-Error (LOE) and Minkowski Distance based Metric (MDM). Table \ref{tab:qualityComparison} demonstrates the effectiveness of our method in achieving a visually pleasing and realistic output. We outperformed other techniques in terms of LOE \cite{DBLP:journals/tip/WangZHL13}, a no-reference image quality metric to assess naturalness in enhanced images. LOE specifically assesses the preservation of lightness order by comparing the enhanced image with the original image, without requiring any additional reference images. A lower LOE score indicates a higher level of preservation of lightness order and, consequently, a more visually pleasing and realistic output. Additionally, we achieved comparable results in MDM, another no-reference quality assessment metric where higher scores indicate higher image quality. 

We observe that our preconditioner can greatly reduce the condition number of the coefficient matrix in \eqref{eq:sol_reflectance} and \eqref{eq:sol_illumination}. Specifically, in some cases, our preconditioner reduces very large condition numbers, which exceeded $5200$, by up to 80\%. 
These findings suggest that our preconditioners can improve the speed of CG execution.

\begin{table}[ht]
\centering
\caption{Computation complexity comparison of LR3M \cite{ren20} and our method based on running time in seconds.}
\label{tab:speedComparison}
\vspace{0.05in}
\begin{footnotesize}
\begin{tabular}{llrr}
\hline
Image        & Resolution       & LR3M \cite{ren20} & proposed  \\ \hline
\texttt{Moonlight}    & $560 \times 420$ & 250.48            & \textbf{25.17} \\
\texttt{Cars}         & $370 \times 415$ & 284.00            & \textbf{16.24} \\
\texttt{Plant}        & $500 \times 375$ & 142.08            & \textbf{21.91} \\
\texttt{Bear}         & $490 \times 365$ & 162.24            & \textbf{20.82} \\
\texttt{Nightfall}    & $690 \times 460$ & 301.75            & \textbf{35.20} \\
\texttt{Street}       & $1035 \times 785$ & 1442.23           & \textbf{88.24} \\
\texttt{Stormtrooper} & $450 \times 450$ & 87.36             & \textbf{21.09} \\
\texttt{Riverside}    & $720 \times 680$ & 366.41            & \textbf{52.82} \\
\texttt{Landmark}     & $540 \times 720$ & 148.01            & \textbf{42.59} \\
\texttt{Man}          & $895 \times 590$ & 246.40            & \textbf{56.47} \\
\texttt{Lamp}         & $450 \times 500$ & 165.40            & \textbf{24.37} \\
\texttt{Wire}         & $325 \times 325$ & 152.29            & \textbf{11.92} \\ \hline
\multicolumn{2}{r}{Average} & 312.39 & \textbf{34.74} \\ \hline
\end{tabular}
\end{footnotesize}
\end{table}

\begin{table*}[ht]
\centering
\caption{Contrast enhancement quality comparison of different methods based on LOE and MDM measures.}
\label{tab:qualityComparison}
\begin{footnotesize}
\begin{tabular}{|
>{\columncolor[HTML]{FFFFFF}}l |
>{\columncolor[HTML]{FFFFFF}}c 
>{\columncolor[HTML]{FFFFFF}}c 
>{\columncolor[HTML]{FFFFFF}}c 
>{\columncolor[HTML]{FFFFFF}}c 
>{\columncolor[HTML]{FFFFFF}}c |
>{\columncolor[HTML]{FFFFFF}}c 
>{\columncolor[HTML]{FFFFFF}}c 
>{\columncolor[HTML]{FFFFFF}}c 
>{\columncolor[HTML]{FFFFFF}}c 
>{\columncolor[HTML]{FFFFFF}}c 
>{\columncolor[HTML]{FFFFFF}}c |}
\hline
\multicolumn{1}{|c|}{\cellcolor[HTML]{FFFFFF}{\color[HTML]{000000} }}                               & \multicolumn{5}{c|}{\cellcolor[HTML]{FFFFFF}{\color[HTML]{000000} LOE \cite{DBLP:journals/tip/WangZHL13}}}                                                                                                                                                                                                                                                                                                                   & \multicolumn{6}{c|}{\cellcolor[HTML]{FFFFFF}{\color[HTML]{000000} MDM \cite{DBLP:journals/tbc/NafchiC18}}}                                                                                                                                                                                                                                                                                                                                                                                                              \\ \hhline{~-----------} 
\multicolumn{1}{|c|}{\multirow{-2}{*}{\cellcolor[HTML]{FFFFFF}{\color[HTML]{000000} Image/Method}}} & \multicolumn{1}{c|}{\cellcolor[HTML]{FFFFFF}{\color[HTML]{000000} M1 \cite{DBLP:conf/cvpr/FuZHZD16}}} & \multicolumn{1}{c|}{\cellcolor[HTML]{FFFFFF}{\color[HTML]{000000} LIME \cite{DBLP:journals/tip/GuoLL17}}}   & \multicolumn{1}{c|}{\cellcolor[HTML]{FFFFFF}{\color[HTML]{000000} JED \cite{DBLP:journals/corr/abs-1804-08468}}}             & \multicolumn{1}{c|}{\cellcolor[HTML]{FFFFFF}{\color[HTML]{000000} LR3M \cite{ren20}}}            & {\color[HTML]{000000} ours}           & \multicolumn{1}{c|}{\cellcolor[HTML]{FFFFFF}{\color[HTML]{000000} Noisy input}} & \multicolumn{1}{c|}{\cellcolor[HTML]{FFFFFF}{\color[HTML]{000000} M1 \cite{DBLP:conf/cvpr/FuZHZD16}}} & \multicolumn{1}{c|}{\cellcolor[HTML]{FFFFFF}{\color[HTML]{000000} LIME \cite{DBLP:journals/tip/GuoLL17}}}   & \multicolumn{1}{c|}{\cellcolor[HTML]{FFFFFF}{\color[HTML]{000000} JED \cite{DBLP:journals/corr/abs-1804-08468}}}             & \multicolumn{1}{c|}{\cellcolor[HTML]{FFFFFF}{\color[HTML]{000000} LR3M \cite{ren20}}}            & {\color[HTML]{000000} ours}           \\ \hline
{\color[HTML]{000000} \texttt{Moonlight}}                                                           & \multicolumn{1}{c|}{\cellcolor[HTML]{FFFFFF}{\color[HTML]{000000} 0.6185}}      & \multicolumn{1}{c|}{\cellcolor[HTML]{FFFFFF}{\color[HTML]{000000} 0.5444}}  & \multicolumn{1}{c|}{\cellcolor[HTML]{FFFFFF}{\color[HTML]{000000} 1.0730}}           & \multicolumn{1}{c|}{\cellcolor[HTML]{FFFFFF}{\color[HTML]{000000} 0.9590}}           & {\color[HTML]{000000} \textbf{0.4817}}         & \multicolumn{1}{c|}{\cellcolor[HTML]{FFFFFF}{\color[HTML]{000000} 0.9594}}      & \multicolumn{1}{c|}{\cellcolor[HTML]{FFFFFF}{\color[HTML]{000000} 0.9522}}      & \multicolumn{1}{c|}{\cellcolor[HTML]{FFFFFF}{\color[HTML]{000000} \textbf{0.9683}}} & \multicolumn{1}{c|}{\cellcolor[HTML]{FFFFFF}{\color[HTML]{000000} 0.9652}}          & \multicolumn{1}{c|}{\cellcolor[HTML]{FFFFFF}{\color[HTML]{000000} 0.9672}}          & {\color[HTML]{000000} 0.9663}         \\
{\color[HTML]{000000} \texttt{Cars}}                                                                & \multicolumn{1}{c|}{\cellcolor[HTML]{FFFFFF}{\color[HTML]{000000} 0.5887}}      & \multicolumn{1}{c|}{\cellcolor[HTML]{FFFFFF}{\color[HTML]{000000} 0.7051}}           & \multicolumn{1}{c|}{\cellcolor[HTML]{FFFFFF}{\color[HTML]{000000} 1.0487}}          & \multicolumn{1}{c|}{\cellcolor[HTML]{FFFFFF}{\color[HTML]{000000} 1.0257}}          & {\color[HTML]{000000} \textbf{0.5479}}         & \multicolumn{1}{c|}{\cellcolor[HTML]{FFFFFF}{\color[HTML]{000000} \textbf{0.9611}}}      & \multicolumn{1}{c|}{\cellcolor[HTML]{FFFFFF}{\color[HTML]{000000} 0.944}}       & \multicolumn{1}{c|}{\cellcolor[HTML]{FFFFFF}{\color[HTML]{000000} 0.9553}}          & \multicolumn{1}{c|}{\cellcolor[HTML]{FFFFFF}{\color[HTML]{000000} 0.9483}}          & \multicolumn{1}{c|}{\cellcolor[HTML]{FFFFFF}{\color[HTML]{000000} 0.9514}}          & {\color[HTML]{000000} 0.9578}         \\
{\color[HTML]{000000} \texttt{Plant}}                                                                        & \multicolumn{1}{c|}{\cellcolor[HTML]{FFFFFF}{\color[HTML]{000000} 0.8414}}      & \multicolumn{1}{c|}{\cellcolor[HTML]{FFFFFF}{\color[HTML]{000000} 0.7030}} & \multicolumn{1}{c|}{\cellcolor[HTML]{FFFFFF}{\color[HTML]{000000} \textbf{0.6998}}}          & \multicolumn{1}{c|}{\cellcolor[HTML]{FFFFFF}{\color[HTML]{000000} 0.8754}}          & {\color[HTML]{000000} 0.8006}         & \multicolumn{1}{c|}{\cellcolor[HTML]{FFFFFF}{\color[HTML]{000000} 0.9633}}               & \multicolumn{1}{c|}{\cellcolor[HTML]{FFFFFF}{\color[HTML]{000000} 0.9582}}      & \multicolumn{1}{c|}{\cellcolor[HTML]{FFFFFF}{\color[HTML]{000000} \textbf{0.9748}}} & \multicolumn{1}{c|}{\cellcolor[HTML]{FFFFFF}{\color[HTML]{000000} 0.9709}}          & \multicolumn{1}{c|}{\cellcolor[HTML]{FFFFFF}{\color[HTML]{000000} 0.9739}}          & {\color[HTML]{000000} 0.9726}         \\
{\color[HTML]{000000} \texttt{Bear}}                                                                         & \multicolumn{1}{c|}{\cellcolor[HTML]{FFFFFF}{\color[HTML]{000000} 0.8517}}      & \multicolumn{1}{c|}{\cellcolor[HTML]{FFFFFF}{\color[HTML]{000000} \textbf{0.6703}}} & \multicolumn{1}{c|}{\cellcolor[HTML]{FFFFFF}{\color[HTML]{000000} 0.9291}}          & \multicolumn{1}{c|}{\cellcolor[HTML]{FFFFFF}{\color[HTML]{000000} 1.3507}}          & {\color[HTML]{000000} 0.7682}         & \multicolumn{1}{c|}{\cellcolor[HTML]{FFFFFF}{\color[HTML]{000000} 0.8384}}               & \multicolumn{1}{c|}{\cellcolor[HTML]{FFFFFF}{\color[HTML]{000000} 0.9076}}      & \multicolumn{1}{c|}{\cellcolor[HTML]{FFFFFF}{\color[HTML]{000000} \textbf{0.976}}}  & \multicolumn{1}{c|}{\cellcolor[HTML]{FFFFFF}{\color[HTML]{000000} 0.975}}           & \multicolumn{1}{c|}{\cellcolor[HTML]{FFFFFF}{\color[HTML]{000000} 0.9756}}          & {\color[HTML]{000000} 0.9716}         \\
{\color[HTML]{000000} \texttt{Nightfall}}                                                                    & \multicolumn{1}{c|}{\cellcolor[HTML]{FFFFFF}{\color[HTML]{000000} 0.8879}}      & \multicolumn{1}{c|}{\cellcolor[HTML]{FFFFFF}{\color[HTML]{000000} 0.7364}}          & \multicolumn{1}{c|}{\cellcolor[HTML]{FFFFFF}{\color[HTML]{000000} 0.4850}}           & \multicolumn{1}{c|}{\cellcolor[HTML]{FFFFFF}{\color[HTML]{000000} 0.7349}} & {\color[HTML]{000000} \textbf{0.4701}}         & \multicolumn{1}{c|}{\cellcolor[HTML]{FFFFFF}{\color[HTML]{000000} 0.9711}}               & \multicolumn{1}{c|}{\cellcolor[HTML]{FFFFFF}{\color[HTML]{000000} 0.9643}}      & \multicolumn{1}{c|}{\cellcolor[HTML]{FFFFFF}{\color[HTML]{000000} 0.9763}}          & \multicolumn{1}{c|}{\cellcolor[HTML]{FFFFFF}{\color[HTML]{000000} 0.9735}}          & \multicolumn{1}{c|}{\cellcolor[HTML]{FFFFFF}{\color[HTML]{000000} \textbf{0.9767}}} & {\color[HTML]{000000} 0.9741}         \\
{\color[HTML]{000000} \texttt{Street}}                                                                       & \multicolumn{1}{c|}{\cellcolor[HTML]{FFFFFF}{\color[HTML]{000000} 0.6415}}      & \multicolumn{1}{c|}{\cellcolor[HTML]{FFFFFF}{\color[HTML]{000000} \textbf{0.5858}}}            & \multicolumn{1}{c|}{\cellcolor[HTML]{FFFFFF}{\color[HTML]{000000} 0.7661}} & \multicolumn{1}{c|}{\cellcolor[HTML]{FFFFFF}{\color[HTML]{000000} 1.2604}}          & {\color[HTML]{000000} 0.7317}         & \multicolumn{1}{c|}{\cellcolor[HTML]{FFFFFF}{\color[HTML]{000000} 0.9706}}               & \multicolumn{1}{c|}{\cellcolor[HTML]{FFFFFF}{\color[HTML]{000000} 0.9682}}      & \multicolumn{1}{c|}{\cellcolor[HTML]{FFFFFF}{\color[HTML]{000000} 0.9762}}          & \multicolumn{1}{c|}{\cellcolor[HTML]{FFFFFF}{\color[HTML]{000000} \textbf{0.9785}}} & \multicolumn{1}{c|}{\cellcolor[HTML]{FFFFFF}{\color[HTML]{000000} 0.9767}}          & {\color[HTML]{000000} 0.9754}         \\
{\color[HTML]{000000} \texttt{Stormtrooper}}                                                                 & \multicolumn{1}{c|}{\cellcolor[HTML]{FFFFFF}{\color[HTML]{000000} 0.9141}}      & \multicolumn{1}{c|}{\cellcolor[HTML]{FFFFFF}{\color[HTML]{000000} 0.7486}}          & \multicolumn{1}{c|}{\cellcolor[HTML]{FFFFFF}{\color[HTML]{000000} 1.2178}}          & \multicolumn{1}{c|}{\cellcolor[HTML]{FFFFFF}{\color[HTML]{000000} 0.8096}}          & {\color[HTML]{000000} \textbf{0.6590}} & \multicolumn{1}{c|}{\cellcolor[HTML]{FFFFFF}{\color[HTML]{000000} 0.9689}}               & \multicolumn{1}{c|}{\cellcolor[HTML]{FFFFFF}{\color[HTML]{000000} 0.9612}}      & \multicolumn{1}{c|}{\cellcolor[HTML]{FFFFFF}{\color[HTML]{000000} 0.9745}}          & \multicolumn{1}{c|}{\cellcolor[HTML]{FFFFFF}{\color[HTML]{000000} 0.9716}}          & \multicolumn{1}{c|}{\cellcolor[HTML]{FFFFFF}{\color[HTML]{000000} 0.9752}}          & {\color[HTML]{000000} \textbf{0.9760}} \\
{\color[HTML]{000000} \texttt{Riverside}}                                                                    & \multicolumn{1}{c|}{\cellcolor[HTML]{FFFFFF}{\color[HTML]{000000} 0.9684}}      & \multicolumn{1}{c|}{\cellcolor[HTML]{FFFFFF}{\color[HTML]{000000} 0.6977}} & \multicolumn{1}{c|}{\cellcolor[HTML]{FFFFFF}{\color[HTML]{000000} 0.7341}}          & \multicolumn{1}{c|}{\cellcolor[HTML]{FFFFFF}{\color[HTML]{000000} 1.1203}}          & {\color[HTML]{000000} \textbf{0.5003}}         & \multicolumn{1}{c|}{\cellcolor[HTML]{FFFFFF}{\color[HTML]{000000} 0.9707}}               & \multicolumn{1}{c|}{\cellcolor[HTML]{FFFFFF}{\color[HTML]{000000} 0.9589}}      & \multicolumn{1}{c|}{\cellcolor[HTML]{FFFFFF}{\color[HTML]{000000} \textbf{0.9758}}} & \multicolumn{1}{c|}{\cellcolor[HTML]{FFFFFF}{\color[HTML]{000000} 0.9721}}          & \multicolumn{1}{c|}{\cellcolor[HTML]{FFFFFF}{\color[HTML]{000000} 0.9707}}          & {\color[HTML]{000000} 0.9739}         \\
{\color[HTML]{000000} \texttt{Landmark}}                                                                     & \multicolumn{1}{c|}{\cellcolor[HTML]{FFFFFF}{\color[HTML]{000000} 0.8279}}      & \multicolumn{1}{c|}{\cellcolor[HTML]{FFFFFF}{\color[HTML]{000000} 0.6800}} & \multicolumn{1}{c|}{\cellcolor[HTML]{FFFFFF}{\color[HTML]{000000} 1.0666}}          & \multicolumn{1}{c|}{\cellcolor[HTML]{FFFFFF}{\color[HTML]{000000} 0.8346}}          & {\color[HTML]{000000} \textbf{0.5080}}          & \multicolumn{1}{c|}{\cellcolor[HTML]{FFFFFF}{\color[HTML]{000000} 0.9571}}               & \multicolumn{1}{c|}{\cellcolor[HTML]{FFFFFF}{\color[HTML]{000000} 0.9435}}      & \multicolumn{1}{c|}{\cellcolor[HTML]{FFFFFF}{\color[HTML]{000000} \textbf{0.9613}}} & \multicolumn{1}{c|}{\cellcolor[HTML]{FFFFFF}{\color[HTML]{000000} 0.9465}}          & \multicolumn{1}{c|}{\cellcolor[HTML]{FFFFFF}{\color[HTML]{000000} 0.9515}}          & {\color[HTML]{000000} 0.9601}         \\
{\color[HTML]{000000} \texttt{Man}}                                                                          & \multicolumn{1}{c|}{\cellcolor[HTML]{FFFFFF}{\color[HTML]{000000} 0.6675}}      & \multicolumn{1}{c|}{\cellcolor[HTML]{FFFFFF}{\color[HTML]{000000} 0.8141}}          & \multicolumn{1}{c|}{\cellcolor[HTML]{FFFFFF}{\color[HTML]{000000} 0.8456}}          & \multicolumn{1}{c|}{\cellcolor[HTML]{FFFFFF}{\color[HTML]{000000} 1.2376}}          & {\color[HTML]{000000} \textbf{0.5209}}         & \multicolumn{1}{c|}{\cellcolor[HTML]{FFFFFF}{\color[HTML]{000000} \textbf{0.9372}}}      & \multicolumn{1}{c|}{\cellcolor[HTML]{FFFFFF}{\color[HTML]{000000} 0.9243}}      & \multicolumn{1}{c|}{\cellcolor[HTML]{FFFFFF}{\color[HTML]{000000} 0.9176}}          & \multicolumn{1}{c|}{\cellcolor[HTML]{FFFFFF}{\color[HTML]{000000} 0.9107}}          & \multicolumn{1}{c|}{\cellcolor[HTML]{FFFFFF}{\color[HTML]{000000} 0.9048}}          & {\color[HTML]{000000} 0.9224}         \\
{\color[HTML]{000000} \texttt{Lamp}}                                                                         & \multicolumn{1}{c|}{\cellcolor[HTML]{FFFFFF}{\color[HTML]{000000} 0.6135}}      & \multicolumn{1}{c|}{\cellcolor[HTML]{FFFFFF}{\color[HTML]{000000} \textbf{0.5644}}} & \multicolumn{1}{c|}{\cellcolor[HTML]{FFFFFF}{\color[HTML]{000000} 1.1641}}          & \multicolumn{1}{c|}{\cellcolor[HTML]{FFFFFF}{\color[HTML]{000000} 1.9776}}          & {\color[HTML]{000000} 0.7960}          & \multicolumn{1}{c|}{\cellcolor[HTML]{FFFFFF}{\color[HTML]{000000} 0.9702}}               & \multicolumn{1}{c|}{\cellcolor[HTML]{FFFFFF}{\color[HTML]{000000} 0.9612}}      & \multicolumn{1}{c|}{\cellcolor[HTML]{FFFFFF}{\color[HTML]{000000} \textbf{0.9733}}} & \multicolumn{1}{c|}{\cellcolor[HTML]{FFFFFF}{\color[HTML]{000000} 0.9721}}          & \multicolumn{1}{c|}{\cellcolor[HTML]{FFFFFF}{\color[HTML]{000000} 0.9709}}          & {\color[HTML]{000000} 0.9705}         \\
{\color[HTML]{000000} \texttt{Wire}}                                                                         & \multicolumn{1}{c|}{\cellcolor[HTML]{FFFFFF}{\color[HTML]{000000} 0.6875}}      & \multicolumn{1}{c|}{\cellcolor[HTML]{FFFFFF}{\color[HTML]{000000} \textbf{0.6515}}} & \multicolumn{1}{c|}{\cellcolor[HTML]{FFFFFF}{\color[HTML]{000000} 1.2575}}          & \multicolumn{1}{c|}{\cellcolor[HTML]{FFFFFF}{\color[HTML]{000000} 1.4944}} & {\color[HTML]{000000} 0.9911}         & \multicolumn{1}{c|}{\cellcolor[HTML]{FFFFFF}{\color[HTML]{000000} 0.9634}}               & \multicolumn{1}{c|}{\cellcolor[HTML]{FFFFFF}{\color[HTML]{000000} 0.9575}}      & \multicolumn{1}{c|}{\cellcolor[HTML]{FFFFFF}{\color[HTML]{000000} \textbf{0.9782}}} & \multicolumn{1}{c|}{\cellcolor[HTML]{FFFFFF}{\color[HTML]{000000} 0.9770}}           & \multicolumn{1}{c|}{\cellcolor[HTML]{FFFFFF}{\color[HTML]{000000} \textbf{0.9782}}} & {\color[HTML]{000000} 0.9729}         \\ \hline
{\color[HTML]{000000} Average}                                                                      & \multicolumn{1}{c|}{\cellcolor[HTML]{FFFFFF}{\color[HTML]{000000} 0.7591}}      & \multicolumn{1}{c|}{\cellcolor[HTML]{FFFFFF}{\color[HTML]{000000} 0.6751}}          & \multicolumn{1}{c|}{\cellcolor[HTML]{FFFFFF}{\color[HTML]{000000} 0.9406}}          & \multicolumn{1}{c|}{\cellcolor[HTML]{FFFFFF}{\color[HTML]{000000} 1.1400}}            & {\color[HTML]{000000} \textbf{0.6480}}          & \multicolumn{1}{c|}{\cellcolor[HTML]{FFFFFF}{\color[HTML]{000000} 0.9526}}               & \multicolumn{1}{c|}{\cellcolor[HTML]{FFFFFF}{\color[HTML]{000000} 0.9501}}      & \multicolumn{1}{c|}{\cellcolor[HTML]{FFFFFF}{\color[HTML]{000000} \textbf{0.9673}}}          & \multicolumn{1}{c|}{\cellcolor[HTML]{FFFFFF}{\color[HTML]{000000} 0.9635}}          & \multicolumn{1}{c|}{\cellcolor[HTML]{FFFFFF}{\color[HTML]{000000} 0.9644}}          & {\color[HTML]{000000} 0.9661}         \\ \hline
\end{tabular}
\end{footnotesize}
\end{table*}

\vspace{-0.1in}
\section{Conclusion}
\label{sec:conclude}
\vspace{-0.05in}
Leveraging on recent advances in GSP, we propose a Retinex-based image denoising / contrast enhancement scheme, where the reflectance and illumination components are optimized alternately using GLR and GGLR for regularization, respectively. 
Both GLR and GGLR are in convenient quadratic form; solutions for reflectance and illumination can be computed as linear systems via conjugate gradient (CG) in roughly linear time. 
We design preconditioners to improve condition numbers of coefficient matrices, speeding up CG.
Experiments show our denoising / contrast enhancement scheme achieved comparable image quality while reducing computation. 




\end{document}